\documentclass[sigconf]{acmart}

\usepackage{algorithm}
\usepackage{algorithmic}
\usepackage{booktabs} 
\usepackage{amsmath}
\usepackage{amssymb}
\usepackage{amsmath}
\usepackage{xcolor}
\usepackage{multirow}
\usepackage{graphicx}
\usepackage{subfigure}
\usepackage{amsfonts}
\usepackage{balance}

\AtBeginDocument{%
  \providecommand\BibTeX{{%
    \normalfont B\kern-0.5em{\scshape i\kern-0.25em b}\kern-0.8em\TeX}}}

\copyrightyear{2020}
\acmYear{2020}
\setcopyright{acmlicensed}\acmConference[SIGIR '20]{Proceedings of the 43rd International ACM SIGIR Conference on Research and Development in Information Retrieval}{July 25--30, 2020}{Virtual Event, China}
\acmBooktitle{Proceedings of the 43rd International ACM SIGIR Conference on Research and Development in Information Retrieval (SIGIR '20), July 25--30, 2020, Virtual Event, China}
\acmPrice{15.00}
\acmDOI{10.1145/3397271.3401145}
\acmISBN{978-1-4503-8016-4/20/07}

\begin{document}
\fancyhead{}

\title{
Learning to Transfer Graph Embeddings for \\ Inductive Graph based Recommendation
}

\author{Le Wu}
\affiliation{Key Laboratory of Knowledge Engineering with Big Data, Hefei University of Technology}
\affiliation{School of Computer Science and Information Engineering, Hefei University of Technology}
\email{lewu.ustc@gmail.com}

\author{Yonghui Yang}
\affiliation{Key Laboratory of Knowledge Engineering with Big Data, Hefei University of Technology}
\affiliation{School of Computer Science and Information Engineering, Hefei University of Technology}
\email{yyh.hfut@gmail.com}

\author{Lei Chen}
\affiliation{Key Laboratory of Knowledge Engineering with Big Data, Hefei University of Technology}
\affiliation{School of Computer Science and Information Engineering, Hefei University of Technology}
\email{chenlei.hfut@gmail.com}

\author{Defu Lian}
\affiliation{University of Science and Technology of China}
\email{liandefu@ustc.edu.cn}

\author{Richang Hong}
\affiliation{Key Laboratory of Knowledge Engineering with Big Data, Hefei University of Technology}
\affiliation{School of Computer Science and Information Engineering, Hefei University of Technology}
\email{hongrc.hfut@gmail.com}

\author{Meng Wang}
\authornote{Corresponding Author}
\affiliation{Key Laboratory of Knowledge Engineering with Big Data, Hefei University of Technology}
\affiliation{School of Computer Science and Information Engineering, Hefei University of Technology}
\email{eric.mengwang@gmail.com}

\begin{abstract}

With the increasing availability of videos, how to edit them and present the most interesting parts to users, i.e., video highlight, has become an urgent need with many broad applications.  As users' visual preferences are subjective and vary from person to person, previous generalized  video highlight extraction models fail to tailor to users' unique preferences. In this paper, we study the problem of personalized video highlight recommendation with rich visual content. By dividing each video into non-overlapping segments, we formulate the problem as a personalized segment recommendation task with many new segments in the test stage. The key challenges of this problem lie in: the \emph{cold-start users} with limited video highlight records in the training data and \emph{new segments} without any user ratings at the test stage.  To tackle these challenges, an intuitive idea is to formulate a user-item interaction graph and perform inductive graph neural network based models for better user and item embedding learning. However, the graph embedding models fail to generalize to unseen items as these models rely on the item content feature and item link information for item embedding calculation. To this end, we propose an inductive \emph{G}raph based \emph{Trans}fer learning framework for personalized video highlight \emph{Rec}ommendation~(\emph{TransGRec}). TransGRec is composed of two parts: a graph neural network  followed by an item embedding transfer network. Specifically, the graph neural network part exploits the higher-order proximity between users and segments to alleviate the user cold-start problem. The transfer network is designed to approximate the learned item embeddings from graph neural networks by taking each item's visual content as input, in order to tackle the new segment problem in the test phase.  We design two detailed implementations of the transfer learning optimization function, and we show how the two parts of TransGRec can be efficiently optimized with different transfer learning optimization functions. Please note that, our proposed framework is generally applicable to any inductive graph based recommendation model to address the new node problem without any link structure. Finally, extensive experimental results on a real-world dataset clearly show the effectiveness of our proposed model.

\end{abstract}



\maketitle

\section{Introduction}
With the increasing availability of camera devices, videos are ubiquitous on entertainment and social networking platforms. As these videos are usually unstructured and long-running, it is  non-trivial to directly browse the interesting and representative parts and share these parts in the social media. Editing such videos into the highlight segments, i.e., the  most interesting or representative parts,  and presenting these parts is a natural choice.  This real-world demand has raised many practical application scenarios, such as increasing user satisfaction for easy GIF creation, sharing and video preview, and boosting the platform prosperity with video promotion and advertising.

In fact, video highlight extraction is closely related to concepts such as video summarization in the computer vision area, which selects representative parts from videos with well-defined objective functions~\cite{CVPR2015video,IJCV2015summarization,CVPR2016highlight}. After that, the same segments are sent to all users. However, users' visual preferences are not universal but vary from person to person. E.g., with a short video that presents the features of a smartphone, the technical fans like the segments that introduce the detailed parameters of the processor, while others prefer to view the photos taken by the smartphone. As a result, instead of presenting the same highlight parts to all users, an ideal video highlight system should suggest and recommend personalized highlight parts to tailor to users' personalized preferences.

Crucially to the success of personalized video highlight recommendation is obtaining users' historical records of the selected highlight parts for personalization. Luckily, many online highlight creation tools, such as \emph{Gifs.com}, have recorded users' manually selection of their liked video intervals, making it possible for the personalization task.  Recently, researchers proposed the problem of personalized highlight recommendation~\cite{MM2018phd}. Intuitively, by dividing each video into a set of non-overlapping segments, the personalized video highlight recommendation task asks: when a user opens a video page, is it possible to automatically suggest (recommend) segments that track her personalized visual preference?  In fact, by drawing analogy between an item from an item set, and a segment from a video, it is natural to deem this task as a recommendation problem.  In the following, for illustration convenience, we do not distinguish the terms of item and segment.

As new videos emerge from time to time, the video highlight recommendation task is quite challenging as many candidate segments at the test time have never appeared in the training data nor rated by any user. Collaborative Filtering~(CF) based recommendation models fail as they rely on the users' historical behaviors to items
and could not generalize to unseen items. Therefore, it is a natural choice to design content based models for personalized video highlight recommendation. With the huge success of deep learning models for image and video processing~\cite{NIPS2012imagenet,ICCV2015learning}, by extracting item semantic features from state-of-the-art deep neural models, most of these models focus on how to align both users and items in a new semantic space by exploiting users' historical behavior data~\cite{NIPS2013music,CVPR2016CDL,MM2018phd}. Some recent works also proposed hybrid recommendation models, where the new item recommendation scenario degenerates to the content based recommendation~\cite{KDD2015CDL,NIPS2017dropoutnet}. All these deep learning based content recommendation models and hybrid models showed better performance to tackle the new item problem. However, in the recommendation process, as each user has very limited historical records, the performances of these models are still far from satisfactory due to the cold-start user problem.

In this paper, we study the problem of personalized video highlight recommendation with two challenges mentioned above: i.e., \emph{cold-start user} problem in the training stage and \emph{new item} without any link structure at the test time. As users naturally form an attributed user-item bipartite graph, an intuitive idea is to perform the graph embedding learning models to model the higher-order graph structure. As such,  the correlations of users' rating records can be exploited for better user and item embedding learning, and the cold-start user problem can be partially alleviated. The graph learning process can be seen as learning a mapping function that takes the item visual feature as well as the item's local link structure as input. However, it fails in the test stage as new items are not rated by any user in the test stage, indicating there is no link structure for item embedding learning. Therefore, a natural idea is that: given the initial item content and the item embedding output by the graph neural network in the training stage, can we design an approximation function to mimic the graph embedding output for the new items in the test stage?

To this end, we propose a general framework: an inductive \emph{G}raph based \emph{Trans}fer learning framework for personalized video highlight \emph{Rec}ommendation~(TransGRec). TransGRec is composed of two parts: a graph neural network and a transfer item embedding learning network. The graph neural network injects the correlations of users' historical item preferences for better user and item embedding learning, and alleviates the user cold-start problem in the training data. The transfer network aims to approximate the item embeddings from graph neural networks by taking an item's visual content as input, such to tackle the missing link structure for the test items. We design two different optimization functions for the transfer network with different assumptions. We show the two parts of TransGRec can be jointly trained in an efficient way. In fact,  our proposed TransGRec framework is generally applicable to inductive graph based recommender models with unseen users or items. Finally, we conduct extensive experimental results on a real-world dataset to show the superiority of our  proposed framework.

\section{Dataset Description} \label{sec:data}
The dataset is based on a popular personalized video highlight website \emph{Gifs.com}, which is introduced by Molino et al.~\cite{MM2018phd}  and publicly available~\footnote{https://github.com/gyglim/personalized-highlights-dataset}.
This website allows users to manually select a video highlight part to create GIFs, which are short and punchy with little space.  When a user selects the time intervals of a video she is interested in, her action is recorded as a quadruple: $<u, v, t\_s, t\_e>$, with $u$ denotes the user, $v$ is the video ID and $t\_s$ denotes the start time of the highlight and $t\_e$ is the end time of the selected highlight part. The original dataset contains about 14,000 users, 119,938 videos and 225,015 annotations.

\begin{small}
\begin{figure} [htb]
  \begin{center}
    \includegraphics[width=80mm]{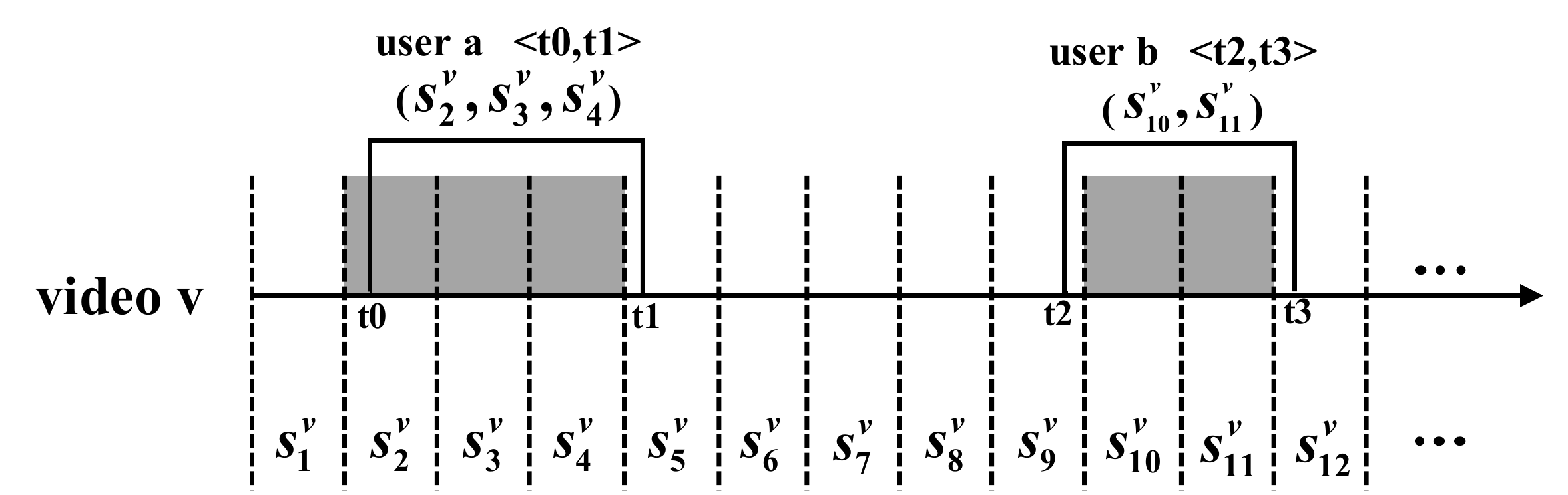}
  \end{center}
    \vspace{-0.5cm}
  \caption{The preprocessing for user-segment records.}\label{fig:video_segment}
   \vspace{-0.2cm}
\end{figure}
\end{small}

\begin{figure} [htb]
  \begin{center}
      \subfigure{\includegraphics[width=40mm]{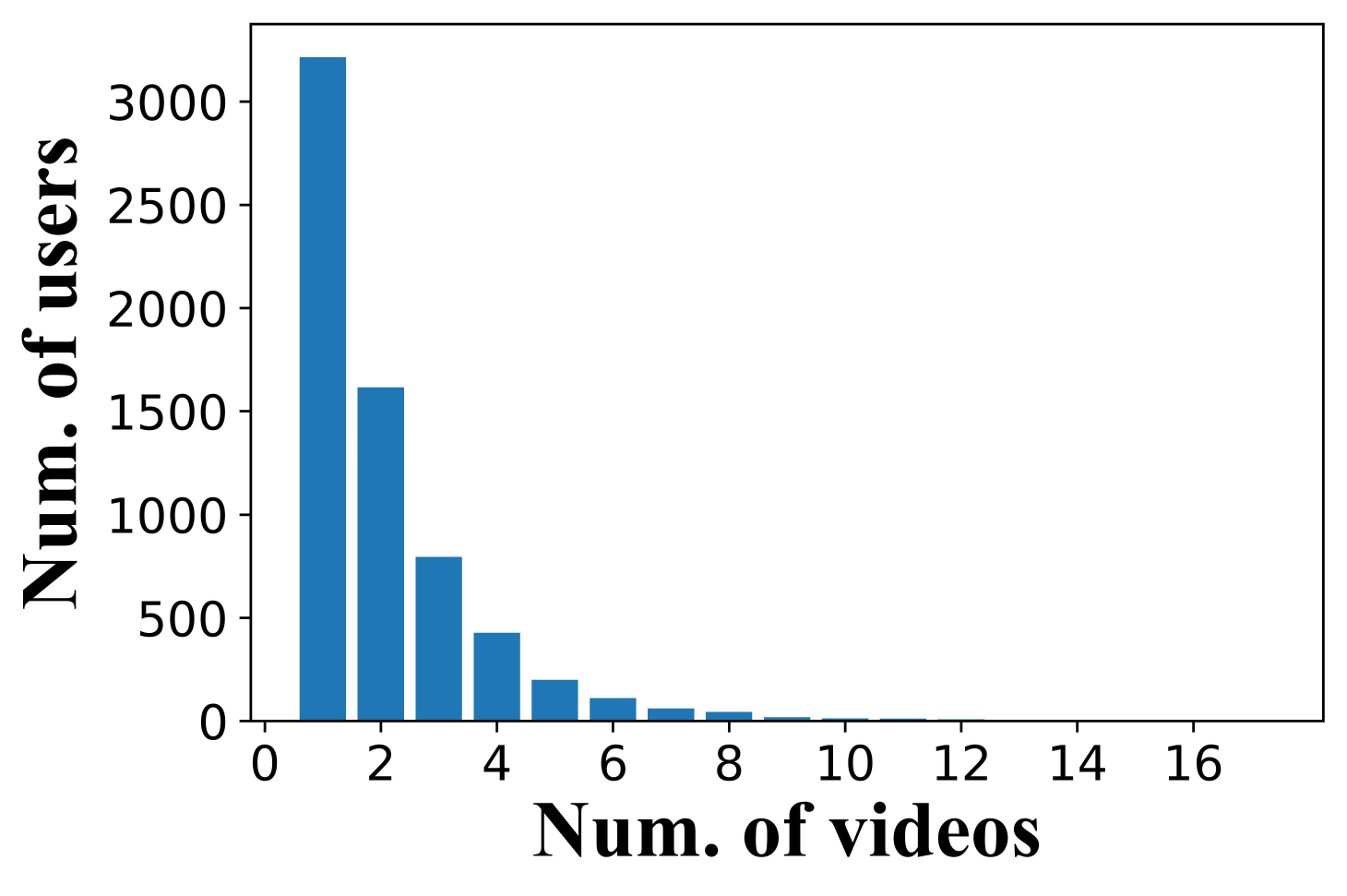}}
      \subfigure{\includegraphics[width=40mm]{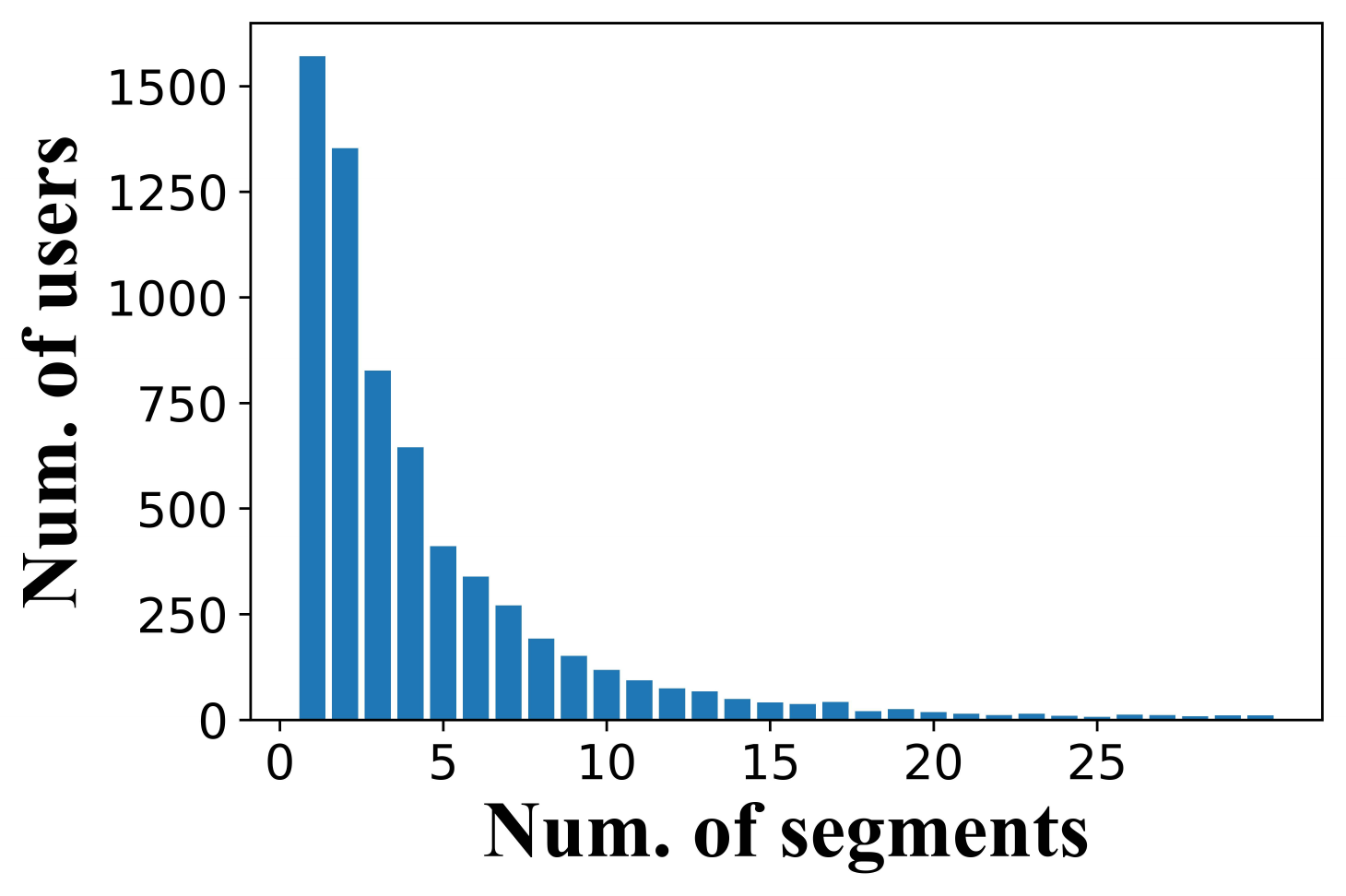}}
  \end{center}%
  \caption{\small{Data distribution: (a)number of rated videos per user; (b)number of rated segments per user. }} \label{fig:data_distribution}
  \vspace{-5pt}
\end{figure}

At the video segment preprocessing step, for each video $v$, similar as many (personalized) video highlight detection models~\cite{CVPR2016highlight,MM2018phd}, we start by dividing each video into a set of non-overlapping segments
$S^v=\{s^v_1, s^v_2,..., s^v_{|v|}\}$. This video segmentation could be implemented by video shot detection based models or simply use the average time split technique~\cite{CBMI2018ridiculously}. As most video highlights are short, we use a simple video segmentation model that equally split 5 seconds as a segment. After that, for each annotation record of user $a$ to video $v$, if the overlap percentage between any segment $s^v_i$~($s^v_i\in S^v$) of this video and her current record is larger than a predefined threshold $\theta$ of the segment, it is considered as user $a$'s positive segments, i.e., $r_{ai}=1$. Without confusion, we would simply replace $s^v_i$ with $i$ to denote a segment. The remaining segments of the video are those that the user has not selected. In this paper, we set the threshold $\theta=50\%$. We show the preprocessing step of two users in Figure~\ref{fig:video_segment}. E.g., for user $a$, her highlight parts span from $s^v_2$ to $s^v_5$, as the liked parts in $s^v_2$ is larger than $\theta$, while the liked parts in $s^v_5$ is smaller than 50\%. Therefore, we set $s^v_2$, $s^v_3$, and $s^v_4$ as the three positive segment records of this user. After data segmentation, we have 6,527 users, 5,137 videos with 55,957 segments. Among the 55,957 segments, 25,777 of them have been liked by users,  with 41,119 user-segment records. For better illustration, we plot the user distribution of rated videos and rated segments in Figure~\ref{fig:data_distribution}. The distributions roughly follow the power law, with many users have very limited rating records.

Let $\mathbf{R}$ denote the user-segment rating matrix, in which $r_{ai}$ is the detailed preference of user $a$ to segment $i$. The personalized video highlight recommendation task asks: with user-segment rating matrix $\mathbf{R}$, for each user $u$ to each test video $v$, our goal is to recommend Top-N ranking list of segments which meet each user's personalized preference.
.

\section{The Proposed Model}
In this section, we introduce a \emph{G}raph based \emph{Trans}fer learning framework for personalized video highlight~\emph{Rec}ommendation~(\emph{TransGRec}). After the data preprocessing, in the training process,  with the user-segment rating matrix $\mathbf{R}$, we construct a user-item bipartite graph $\mathcal{G}=<\mathcal{U}\cup\mathcal{I}, \mathbf{R}>$, with $\mathcal{U}$ denotes the user set with $M$ users~({\small$|\mathcal{U}|=M$}) and $\mathcal{I}$~({\small$|\mathcal{I}|=N$}) is the item~(segment) set extracted from all the training data. $\mathbf{R}=[r_{ai}]_{M\times N}$ is the edge set, with $r_{ai}=1$ denotes user $a$ shows preference for segment $i$. As videos change quickly in the real world, in the test stage, most test videos are new and have not been rated by users. In other words, these test items neither appear in the training data, nor have been rated by any users.

Given the characteristics of the problem, as shown in Figure~\ref{fig:framework}, there are two key parts of the proposed framework. First, by turning users' historical records into a graph structure, TransGRec represents the user-segment rating behavior as an attributed graph, and uses Graph Neural Networks~(GNN) for user and item representation learning. Therefore, the higher-order graph structure is leveraged in the user and item embedding learning process, which could alleviate the cold-start problem. Second, most test videos are new and never appear in the training data, neither do they have any rating records. Under such a situation, graph neural network based inductive learning models fail as each item relies on its content and the local link structure for item embedding learning. To tackle the new item situation without edge in the test stage, we propose a transfer learning model~($T$) that learns to transform each item's visual input into an approximated embedding in the GNN output space. The goodness of the transfer network is measured by comparing the results of the approximated item embedding, and the real item embedding from the output of the graph neural network in the training data. Therefore, in the test stage, for each item that has not appeared in the training data, the transfer network could be applied to approximate each item's embedding in the inductive setting without any link structure. Then, the predicted rating is learned by comparing the similarity between the user-item pair in the final embedding space.

Given the two parts of the proposed TransGRec structure, the overall loss function is naturally defined as combining the loss functions of the two parts as:

\begin{small}
\vspace{-0.2cm}
\begin{equation} \label{eq:overall_loss}
\mathcal{L}=\mathcal{L}_{GNN}+\lambda\mathcal{L}_T,
\end{equation}
\vspace{-0.2cm}
\end{small}

\noindent where~preference prediction loss~($\mathcal{L}_{GNN}$) is the classical rating based optimization loss under GNN modeling, and transfer loss~($\mathcal{L}_T$) denotes the loss for the transfer network $T$. $\lambda$  is a balance parameter between these two loss functions.

\begin{small}
\begin{figure*} [htb]
  \begin{center}
    \includegraphics[width=160mm]{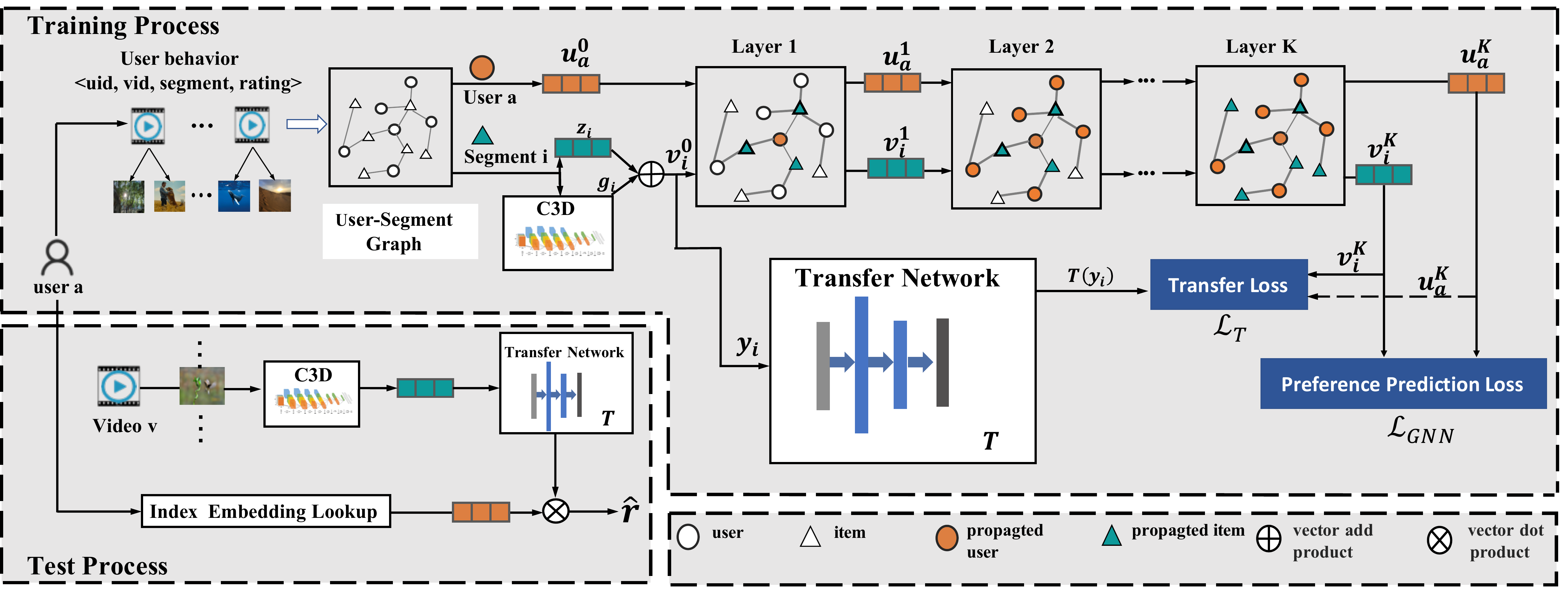}
  \end{center}
    \vspace{-0.5cm}
  \caption{The overall framework of TransGRec.}\label{fig:framework}
   \vspace{-0.3cm}
\end{figure*}
\end{small}

\subsection{Graph Neural Network for Embedding Learning}
The graph neural network is mainly composed of four parts: the initial embedding layer, the item embedding fusion layer, the recursive user and item propagation layers, and the prediction layer.

\textbf{Initial embedding layer.} Similar to many embedding based recommendation models, we use
{\small$\mathbf{X}\in\mathbb{R}^{D\times M}$} and {\small$\mathbf{Z}\in\mathbb{R}^{D\times N}$} to denote the free embedding matrices of users and items. For each user, as we do not have her profile or related metadata, each user $a$'s initial embedding is denoted as $\mathbf{x}_a$, which is the $a$-th column of the user free embedding matrix {\small$\mathbf{X}\in\mathbb{R}^{D\times M}$}. For each item $i$, its free embedding is the $i$-th column of the item free embedding matrix {\small$\mathbf{Z}\in\mathbb{R}^{D\times N}$}, i.e., $\mathbf{z}_i$.

\textbf{Item embedding fusion layer.} The item embedding fusion layer fuses the free item embedding and its visual embedding. For each segment that appears in a video, we leverage video representation models for item semantic embedding. Given each item of a video segment, similar to many popular visual-based video processing approaches~\cite{ji20123d,ICCV2015learning}, we select  Convolutional 3D  Networks~(C3D) pretrained on the Sports-1M~\cite{CVPR2014sports1M} dataset as the video feature extractor. Specifically, we use the output of first connected layer (fc1) of C3D as the visual feature {\small$\mathbf{f}_i\in\mathbb{R}^{4096 \times 1}$} of each item $i$. Then, we use a dimension reduction matrix $\mathbf{W}^0\in \mathbb{R}^{D\times 4096}$ to reduce the original visual embedding into a low dimension space as:

\begin{small}
\vspace{-5pt}
\begin{equation}
\mathbf{g}_i=\mathbf{W}^0\mathbf{f}_i \label{eq:item_content}.
\end{equation}
\vspace{-5pt}
\end{small}

Then, for each item  $i$, we could get its fused item embedding as a combination of the free embedding and the visual content embeddding:

\begin{small}
\vspace{-5pt}
\begin{equation}
\mathbf{y}_i= \mathbf{g}_i+\mathbf{z}_i= \mathbf{W}^0\mathbf{f}_i+\mathbf{z}_i\label{eq:item_gcn_v0}.
\end{equation}
\vspace{-5pt}
\end{small}

Please note that, there are different kinds of methods that fuse the two parts of item representations, such as concatenation or addition. In this paper, we use the addition operation, as we find the addition operation is more stable in the experiments, and usually achieves better performance.

\textbf{Embedding propagation layers.} This part stacks multiple layers to propagate user and item embeddings of the graph, such that the higher-order graph structure is modeled to refine user and item embeddings. Let $\mathbf{u}^{k}_a$ denote user $a$'s embedding and $\mathbf{v}^{k}_i$ as item $i$'s embedding at the $k$-th propagation layer.
As the output of the initial embedding layer is directly sent to the propagation layers,  we have $\mathbf{u}_a^0=\mathbf{x}_a$ and $\mathbf{v}_i^{0}=\mathbf{y}_i$ with $k=0$.

By iteratively feeding the output of all the nodes' embeddings in the graph from $k$-th propagation layer, each user $a$'s updated embedding $\mathbf{u}^{(k+1)}_a$ at $(k+1)$-th is composed of two steps:  a pooling operation that aggregates all the connected neighbors' embeddings at $k$-th layer into a fixed-length vector representation $\mathbf{u}_{Ra}^{k+1}$, followed by an update step that combines the neighbors' representations and her own embedding at $k$-th layer.  Mathematically, these two steps could be defined as:

\begin{small}
\vspace{-5pt}
\begin{flalign}
\mathbf{u}_{Ra}^{k+1}=Pool(\mathbf{v}_j^{k}|j\in R_a), \label{eq:gcn_su}\\
\vspace{-0.2cm}
\mathbf{u}_a^{k+1}=\sigma(\mathbf{W}_u^{k+1}\times(\mathbf{u}^k_a+\mathbf{u}_{Ra}^{k+1})),  \label{eq:gcn_u}
\end{flalign}
\vspace{-5pt}
\end{small}

\noindent where {\small$R_a=\{i|R_{ai}\!=\!1\}, R_a\subseteq V$} is the item set that user $a$ interacts with. {\small$\mathbf{W}_u^{k+1}\in\mathbb{R}^{D\times D}$} is the transformation matrix in $(k+1)$-th layer, and $\sigma(x)$ is an activation function. In Eq.~\eqref{eq:gcn_su}, the pooling operation in the item neighbor aggregation step is quite flexible, which could be defined as an average pooling that takes the mean of all item neighbors' embedding vectors, or the max pooling that selects the maximum value in each dimension from the embeddings of all the item neighbor set. In the update step of Eq.\eqref{eq:gcn_u}, the pooling vector is first added with the user representation in the previous layer, and then a transformation to get the updated user embedding at $(k+1)$-th layer. Please note that, we have also tried the concatenation operation, and find the addition also performs better.

Given the user embedding process, similarly, let {\small$R_i=\{a|R_{ai}\!=\!1\}, R_i\subseteq U$} denote the user set that shows preferences to item $i$, we could update each item $i$'s embedding $\mathbf{v}^k_{i}$ at the $k$-th layer to $\mathbf{v}^{(k+1)}_i$ at $(k+1)$-th layer as:

\begin{small}
\vspace{-5pt}
\begin{flalign}
\mathbf{v}_{Ri}^{k+1}=Pool(\mathbf{u}_a^{k}|a\in R_i), \label{eq:gcn_si}\\
\vspace{-0.2cm}
\mathbf{v}_i^{k+1}=\sigma(\mathbf{W}_v^{k+1}\times(\mathbf{v}^k_i+ \mathbf{v}_{Ri}^{k+1})),  \label{eq:gcn_i}
\end{flalign}
\vspace{-5pt}
\end{small}

\textbf{Prediction layer.}
After the embedding propagation layers reaches a defined depth $K$, we obtain the user embedding $\mathbf{u}_a^{K}$ and the item embedding $\mathbf{v}_i^{K}$ at the $K$-th layer for final user and item embedding:

\begin{small}
\vspace{-5pt}
\begin{flalign}
\mathbf{u}_a= \mathbf{u}_a^{K}, \label{eq:final_userembed}\\
\vspace{-0.2cm}
\mathbf{v}_i= \mathbf{v}_i^{K}. \label{eq:final_userembed}
\end{flalign}
\vspace{-5pt}
\end{small}

Then, we could predict each user $a$'s rating to item $i$ as the inner product between the user embedding $\mathbf{u}_a$ and item embedding $\mathbf{v}_i$:

\vspace{-0.2cm}
\begin{equation}
\hat{r}_{ai}= \mathbf{u}^T_a\mathbf{v}_i.
\end{equation}
\vspace{-0.2cm}

Please note that, in the graph neural network part for embedding learning, distinct from classical inductive graph embedding models that take item content as input, we additional add the free item embedding into embedding propagation layers.  Adding the free item embedding could allow the graph neural network to learn the collaborative information that is not reflected in the content, which usually shows better performance for many recommendation tasks. Therefore, the final item embedding output from the graph neural network is a hybrid item representation.  Naturally, the above graph neural network part is not inductive and could not generalize to unseen items in the test stage. We would illustrate how to achieve the inductive ability by transfer network introduced in the following subsection.

\subsection{Transfer Network for Inductive Item Embedding Learning}~\label{subsec:transfer}

As new videos evolve from time to time, in the test stage, most candidate items~(segments)  have never appeared in the training data. More importantly, most of these new items have not been rated by any users, i.e., most test items are isolated nodes and are not connected to any user in the training data. As  each new item's free embedding is not available, as shown in Eq.\eqref{eq:item_gcn_v0} and Eq.\eqref{eq:gcn_si}, the item embedding fusion layer and the propagation layers fail. To tackle the new item in the test stage, a natural idea is to directly send each candidate item $i$'s content features into a transfer network, and outputs its final embedding $\mathbf{v}_i$ with the learned parameters. Under such setting, we design a transfer network $T$ to transform each item $i$'s embedding in the content space $\mathbf{f}_i$ to approximate its embedding in the final embedding space $\mathbf{v}_i$.

In fact, the mapping function from each item's visual input representation to its final input learned from GNN is complex. To tackle the complex transformation, we choose a Multi-Layer Perception~(MLP) to implement the transfer network, as MLPs are demonstrated to be able to approximate any complex function~\cite{hornik1989multilayer}. Therefore, given all pairs of $[\mathbf{f}_i, \mathbf{v}_i]_{i=1}^N$ in the training data as labeled data, the transfer network $T$ with a MLP learns to approximate the output embeding as:

\begin{small}
\vspace{-0.3cm}
  \begin{equation}\label{eq:transnet}
   \hat{\mathbf{v}}_i=T(y_i)=h^L(... h^1(\mathbf{y}_i)),
  \end{equation}
\vspace{-0.1cm}
\end{small}

\noindent where $h^l(x)=f(\mathbf{P}^l h^{(l-1)}(x))$ is a function that takes the output of the $(l-1)$-th layer as input. $\mathbf{P}^l$ is the parameter at the l-th layer. Note that as each item's collaborative embedding should be consistent with the range of the output in the~$GNN$, we set $f(x)$ same as~$\sigma$ of Eq.~\eqref{eq:gcn_i}.

\textbf{Euclidean distance based loss.}
A naive idea of defining  the transfer network  $T$ is to compare the learned transferred embeddigns with output embeddings from GNN in the Euclidean space as:

\begin{small}
\vspace{-0.2cm}
\begin{equation} \label{eq:loss_t_dis}
\mathcal{L}_1 =\sum_{i=1}^I ||\hat{\mathbf{v}}_i-\mathbf{v}_i||^2_F,
\end{equation}
\vspace{-0.2cm}
\end{small}

In the above transfer network $T$, let~$\Theta_T=[\mathbf{P}^l]_{k=1}^L$ denote the parameter set. The above Euclidean space assumes that conditional likelihood of the approximated item embedding matrix $\hat{\mathbf{V}}$ learned from the transfer network $T$ follows a Gaussian distribution as: $p(\hat{\mathbf{V}}|\mathbf{V})\sim\mathcal{N}(mean, \sigma)$, with the mean value is the item embedding learned from the graph neural network $GNN$. As such, maximizing the log likelihood of the approximated item embedding is equivalent to minimizing the Euclidean distance as shown in Eq.\eqref{eq:loss_t_dis}.

\textbf{Adversarial loss.}
The above Euclidean distance~(Eq.~\ref{eq:loss_t_dis}) fails when $p(\hat{\mathbf{V}}|\mathbf{V})$ is complex, e.g., this distribution has multiple peak points. Simply reducing this likelihood function to a Guassian distribution would fail. Therefore, in order to model the complex conditional distribution, we propose to introduce an adversarial loss based on Generative Adversarial Networks~\cite{goodfellow2014generative}. Besides the transfer network $T$ that generates ~(fake) approximated item embeddings, a discriminator $D$ parameterized by $\Theta_D$ is introduced to distinguish whether the item embedding is real or fake. Specifically, the ``real'' labels are assigned to  the item embeddings that are learned from the output of GNN, while ``fake'' samples are those approximated item embeddings learned by the transfer network $T$.
The adversarial loss is defined to let the discriminator $D$ and the transfer network $T$ to play the following two-player minimax game as:

\begin{flalign}\label{eq:loss_t_adv}
\arg\max_{\theta_T}\min_{\theta_D}\mathcal{L}_{2}=&-\sum_{i=1}^N\sum_{u=1}^M [log D(\mathbf{v}_i,\mathbf{u}_a,r_{ai})+\nonumber \\
&log (1-D(T(\mathbf{y}_i),\mathbf{u}_a,r_{ai}))],
\end{flalign}

In the above optimization function, we also feed $\mathbf{u}_a$~and~$r_{ai}$ into the discriminator, as introducing these conditional or latent information would make the adversarial training process easier~\cite{chen2016infogan,antipov2017face}. Since $T$ is implemented with MLPs in Eq.\eqref{eq:transnet}, we also use another MLPs to constitute the discriminator $D$, with $\Theta_D=[\mathbf{Q}^l]_{l=1}^L$ is the parameter set in this discriminator.

\subsection{Model Optimization}
As the implicit preference is more common in real-world recommender systems, without loss of generality, we use Bayesian Personalized Ranking~(BPR) as the preference loss function in GNN~\cite{BPR2009}:

\begin{small}
\vspace{-0.3cm}
  \begin{equation}\label{eq:loss_r}
  \min\limits_{\Theta_{GNN}} \mathcal{L}_{GNN}=\sum_{a=1}^M\sum\limits_{(i,j)\in D_a } -\mathbf{ln}~s(\hat{r}_{ai}-\hat{r}_{aj}) +\lambda(||\mathbf{X}||^2+||\mathbf{Z}||^2),
  \end{equation}
\vspace{-0.1cm}
\end{small}

\noindent where $s(x)$ is a sigmoid function, {\small $\Theta_{GNN}\!=\![\mathbf{X},\mathbf{Z},\Theta_2=[\mathbf{W}^k]_{k=0}^K] ]$} is the parameter set in the graph neural network,  and  $\lambda$ is a regularization parameter. {\small$D_a=\{(i,j)|i\in R_a\!\wedge\!j\not\in R_a\}$} denotes the pairwise training data for $a$, with {\small$R_a$} represents the item set that $a$ positively shows feedback and $j$ belongs to the negative itemset. For each user of a positive segment $i$, it is associated with a corresponding video. Therefore, all the segments that have not been rated by the user are considered as candidate negative items.

\textbf{Overall optimization function.}
Given the overall optimization function in Eq.\eqref{eq:overall_loss}, with the detailed GNN based loss~(Eq.\eqref{eq:loss_r}) and two detailed optimization functions in $T$, we can get two kinds of overall optimization functions.

\textbf{1) Overall optimization function with Euclidean loss.}
For convenience, We call this proposed model as TransGRec-E~(Euclidean). By combining the Euclidean loss in Eq.\eqref{eq:loss_t_dis}, the  overall loss function is defined as:

\begin{flalign}\label{eq:overall_loss_dis}
 \arg\min_{[\Theta_T,\Theta_{GNN}]}\mathcal{L}= &\sum_{a=1}^M\sum\limits_{(i,j)\in D_a }-\mathbf{ln}~s(\hat{r}_{ai}-\hat{r}_{aj}) +\nonumber  \\
 &\lambda(||\mathbf{X}||^2+||\mathbf{Z}||^2)+\sum_{i=1}^I ||T(\mathbf{y}_i)-\mathbf{v}_i||^2_F
\end{flalign}

As all parameters in the optimization function are differentiable, we could use gradient descent algorithms to optimization.

\textbf{2) Overall optimization function with adversarial loss.} For convenience, we call this proposed model as TransGRec-A~(Adversarial). With the adversarial loss in Eq.\eqref{eq:loss_t_adv}, there are three sets of parameters as: {\small$\Theta=[\Theta_{GNN},\Theta_T, \Theta_D]$}. Since the overall optimization function involves both maximization and minimization with regard to different parameter sets. We use alternating update step as:

\begin{itemize}

  \item Fix $\Theta_{GNN}$ and $\Theta_D$, update $\Theta_T$ as:
    \begin{equation}\label{eq:overall_loss_t}
    \begin{split}
    \arg\min_{\Theta_T}\sum_{i=1}^N\sum_{a=1}^M [log (1-D(T(\mathbf{y}_i)),\mathbf{u}_a,r_{ai})].
    \end{split}
    \end{equation}

  \item Fix $\Theta_{GNN}$ and $\Theta_T$, update $\Theta_D$ as:

    \begin{equation}\label{eq:overall_loss_d}
    \begin{split}
    \arg\min_{\Theta_D} \sum_{i=1}^N\sum_{a=1}^M -[log D(\mathbf{v}_i,\mathbf{u}_a,r_{ai})+
    log (1-D(T(\mathbf{y}_i)),\mathbf{u}_a,r_{ai})].
    \end{split}
    \end{equation}

\end{itemize}

In practice, we alternate the process of updating $\Theta_{GNN}$, $\Theta_{T}$ and $\Theta_{D}$, and stops when the loss function converges.

After the training process is finished, the personalized video recommendation process is very easy. The overall flowchart of the test phase is shown at the bottom part of Figure ~\ref{fig:framework}. For each user $a$, we can index the user's final embedding $\mathbf{u}_a$ from the learned user embedding matrix {\small$\mathbf{U}$}. For each candidate test segment $i$, we first take the item feature $\mathbf{f}_i$ as input and get the low dimension visual embedding~$\mathbf{g}_i$. Then we can get the approximated item embedding from the transfer network as: $\hat{\mathbf{v}}_i=T(\mathbf{g}_i)$. In summary, the predicted preference of each user-item pair $(a,i)$ can be calculated as:

\begin{equation}\label{eq:pred_test}
\hat{r}_{ai}=u_a^T\hat{\mathbf{v}}_i
\end{equation}

\subsection{Discussion}

\textbf{Connections with related learning models.}
TransGRec framework is a concrete application of transfer learning~\cite{tkde2009survey}.
We formulate the video highlight recommendation problem as an inductive representation learning on graph neural networks. In order to generalize to unseen nodes in test stage, we transfer the embeddings learned in the source network~(i.e., GNN) of the training data to learn an approximated item embedding in the target network. In the review based recommendation problem, some researcheres have applied the transfer network to approximate each unseen user-item pair review with the available information in the training data~\cite{recsys2017transnets,www2020dualnet}. We differ greatly from these works from the problem definition, the proposed model and the application. These review based recommendation models are transductive learning problems with missing values in the test stage, and do not rely on the graph neural network based approaches. By using MLPs to approximate the graph neural network structure, our work also seems to be correlated with knowledge distillation~\cite{hinton2015distilling,iclr2016unifying}.  The graph neural network can be regarded as the complex teacher network, while the transfer network resembles a student network that distills the knowledge learned from the teacher network.

\textbf{Generalization of the proposed model.}
In fact, TransGRec could be seen as an inductive graph based  hybrid recommendation framework. With user-item bipartite graph, the item embedding is fused by the free collaborative embedding and the content embedding extracted from video representation learning models. Therefore, the graph neural network could better refine user and item embeddings by  preserving the higher-order local structure of this user with recursive feature propagation in this graph. In such a way, we could better represent users and items to alleviate the data sparsity issue. The transfer network learns to transfer the available content representation to the final graph embedding space. Therefore,  the proposed framework is generally applicable to many inductive recommendation scenarios with new multimedia items or cold-start users. E.g., in the news recommendation scenario,  news articles are highly time-sensitive with many news come from time to time, our proposed framework could well utilize both the article content and the collaborative signals in users'  previous behaviors for better user and news representation learning.

\begin{table}[t]
  \centering
  \setlength{\belowcaptionskip}{3pt}
  \caption{The statistics of the dataset.}\label{tab:statistic}
  \vspace{-5pt}
  \begin{tabular}{|c|c|c|}
    \hline
    Users & Segments &Segments in the graph\\
    \hline
    6,527& 55,957 &25,777\\
    \hline
    Training records& Validation records & Test records\\
    \hline
    34,563 &3,840&2,716\\
    \hline
  \end{tabular}
  \vspace{-15pt}
\end{table} 

\section{Experiments}

\subsection{Experimental Settings}
\textbf{Experimental setup.} As each highlight record is associated with a detailed time,  with the preprocessed dataset introduced before, for each user that has at least five video highlight records, we select the last highlight video of each user for model evaluation, which leads to 2716 user-segment records of the test data. Then, we randomly select 10\% from the remaining data as the validation set for model tuning. We show the data statistics of the dataset in Table~\ref{tab:statistic}.

In model evaluation stage, for each user-segment record,  it is associated with a video. We use the unobserved segments of the corresponding video that have not been selected by the user as candidate negative items for recommendation. We use two kinds of metrics from the video highlight domain and the recommendation domain for evaluation. First, as the personalized video highlight recommendation is correlated to video highlight extraction, we adopt  two  widely used video detection metrics: Mean Average Precision~(MAP) and Normalized Meaningful Summary Duration~(NMSD) to evaluate the prediction performance on the video level~\cite{CVPR2016video2gif,MM2018phd}. Specifically, MAP describes the average precision of the ranking of all segments of a video, and the larger value means the better performance. NMSD rates how much of the segments have been watched before the majority of the ground truth segments were shown, so the smaller value means the better performance. Besides, our task is also a Top-N recommendation task, so we use three popular ranking metrics for evaluation from different perspectives:  Hit Ratio~(HR), Recall~\cite{SIGIR2018prp}, and Normalized Discounted Cumulative Gain~(NDCG)~\cite{SIGIR2018attentive,IJCAI2019personalized}.  HR measures the number of items in the Top-N ranking list that the user likes. Recall measures the number of items that the user likes in the test data that has been successfully predicted in the Top-N ranking list. And NDCG considers the hit positions of the items and gives a higher scores for the hit items in the topper positions.
.

\begin{table}[t]
  \begin{small}
    \centering
    \setlength{\belowcaptionskip}{5pt} %
    \caption{Overall performance comparison {$\uparrow$} means the larger value, the better performance; {$\downarrow$} means the smaller value, the better performance).}
    \vspace{-5pt}
    \label{tab:overall comparision}
      \scalebox{0.9}
    {\begin{tabular}{|c|c|c|c|c|c|}
        \hline
        Models&MAP$\uparrow$&NMSD$\downarrow$&HR@5$\uparrow$&NDCG@5$\uparrow$&Recall@5$\uparrow$\\
      \hline
      Video2GIF&0.2075&0.4288&0.1993&0.1651&0.1798\\
      \hline
      SVM-D&0.2185&0.4180&0.2191&0.1772&0.1991\\
      \hline
      PHD-GIFs&0.2170&0.4419&0.2228&0.1781&0.2028\\
      \hline
      DropoutNet&0.2604&0.3886&0.2569&0.2162&0.2353 \\
      \hline
        CDL&0.2706&0.3806&0.2729&0.2304&0.2540\\
        \hline
            LapReg& 0.2828& 0.3905& 0.2794&0.2360&0.2598\\
        \hline
        \emph{\textbf{TransGRec-E}}&0.3113&\textbf{0.3702}&{0.3066}&{0.2687}&{0.2873}\\
        \hline
        \emph{\textbf{TransGRec-A}}&\textbf{0.3174}&0.3769&\textbf{0.3153}&\textbf{0.2779}&\textbf{0.2940}\\
        \hline
      \end{tabular}}
  \end{small}
\end{table}

\begin{table*}
  \begin{small}
    \centering
    \setlength{\belowcaptionskip}{2pt}
    \caption{Recommendation metrics comparisons of different Top-N values.}\label{tab:hr_ndcg_recall_topk}
    \vspace{-5pt}
    \resizebox{\textwidth}{!}{
    \begin{tabular}{|c|c|c|c|c|c|c|c|c|c|c|c|c|c|c|c|}
      \hline
      \multirow{2}{*}{Models} &
      \multicolumn{5}{|c|}{HR@N $\uparrow$} &\multicolumn{5}{|c|}{Recall@N $\uparrow$} &\multicolumn{5}{|c|}{NDCG@N$\uparrow$} \\
      \cline{2-16}
      &N=1&N=2&N=3&N=4&N=5&N=1&N=2&N=3&N=4&N=5&N=1&N=2&N=3&N=4&N=5\\
      \hline
      Video2GIF&0.1477&0.1319&0.1445&0.1702&0.1993&0.0498&0.0795&0.1112&0.1465&0.1798&0.1477&0.1348&0.1398&0.1521&0.1651\\
      \hline
            SVM-D&0.1371&0.1340&0.1656&0.1934&0.2191&0.0513&0.0863&0.1311&0.1685&0.1991&0.1371&0.1332&0.1505&0.1640&0.1772\\
      \hline
      PHD-GIFs&0.1435&0.1445&0.1586&0.1966&0.2228&0.0478&0.0888&0.1246&0.1723&0.2028&0.1435&0.1428&0.1480&0.1661&0.1781\\
      \hline
            DropoutNet&0.1751&0.1804&0.2057&0.2249&0.2569&0.0617&0.1191&0.1641&0.1958&0.2353&0.1751&0.1774&0.1920&0.2015&0.2162\\
      \hline
      CDL&0.1857&0.1941&0.2278&0.2514&0.2729&0.0714&0.1340&0.1890&0.2254&0.2540&0.1857&0.1901&0.2086&0.2208&0.2304\\
      \hline
       LapReg&0.1793 &0.1962&0.2215&0.2472&0.2794&0.0746&0.1399&0.1833&0.2203&0.2598&0.1793&0.1912&0.2072&0.2208&0.2360\\
      \hline
      \emph{\textbf{TransGRec-E}}&0.2215&0.2373&0.2595&0.2815&{0.3066}&{0.0948}&0.1712&0.2171&0.2548&{0.2873}&0.2215&0.2323&0.2453&0.2563&{0.2687}\\
      \hline
      \emph{\textbf{TransGRec-A}}&\textbf{0.2363}&\textbf{0.2416}&\textbf{0.2634}&\textbf{0.2915}&\textbf{0.3153}&\textbf{0.1045}&\textbf{0.1767}&\textbf{0.2220}&\textbf{0.2620}&\textbf{0.2940}&\textbf{0.2363}&\textbf{0.2395}&\textbf{0.2519}&\textbf{0.2668}&\textbf{0.2779}\\
      \hline
    \end{tabular}}
    \vspace{-5pt}
  \end{small}
\end{table*}

\textbf{Baselines.} We compare our proposed model with the following baselines: Video2GIF~\cite{CVPR2016video2gif}, SVM-D~\cite{ECCV2014ranksvm}, PHD-GIFs~\cite{MM2018phd},
DropoutNet~\cite{NIPS2017dropoutnet}, CDL~\cite{CVPR2016CDL}, and LapReg~\cite{NIPS2017inductive}. Specifically, Video2GIF is a state-of-the-art model for general video highlight detection, which exploits the different visual representations of positive items and negative items for modeling. SVM-D and PHD-GIFs are two popular models designed for content based recommendation, with the user embedding is learned from their interacted items. DropoutNet and CDL are content enhanced model that could tackle the new item problem. Specifically,
DropoutNet deals with the new user and new item  with user dropout and item dropout in the training process~\cite{NIPS2017dropoutnet}. CDL learns the free user content embedding, and the item content embedding in a low latent space given the available training data~\cite{CVPR2016CDL}. LapReg is designed for semi-supervised tasks, with the prediction function similar to the supervised learning models, and it has an additional Laplacian regularization term to capture the correlations of predictions in the graph~\cite{JMLR2006manifold}. In practice,
for fair comparison of LapReg, we choose the best prediction model in the remaining baselines as the prediction function in LapReg, and the Laplacian regularization is performed in the user-item bipartite graph. Therefore, these baselines include classical supervised models, deep learning based supervised models, content enhanced recommendation models that can tackle the new item problem, and the graph based  semi-supervised models. Please note that, we do not introduce any  inductive graph neural network based recommendation models for recommendation, as these models need the unseen nodes to have several links. However, in the video highlight recommendation, most new segments have never been rated by any user. Therefore, these inductive graph neural network based models fail.

\textbf{Parameter setting.}
For baselines of Video2GIF and PHD-GIFs, we use the same settings as the original papers with two hidden layer dimensions of 512 and 128~\cite{CVPR2016video2gif,MM2018phd}. For all embedding based models (SVM-D, CDL, DropoutNet, TransGRec), we initialize the embedding matrix with a Gaussian distribution with a mean of 0 and variance of 0.01.
In the model training process~(Eq.\eqref{eq:loss_r}), as the unobserved possible negative samples are much more than positive samples, similar to many implicit feedback based recommendation models~\cite{BPR2009,WWW2017generic}, we use negative sampling technique at each training iteration. Specifically, for each positive pair of user $a$ to item $i$ that comes from video $v$,  we choose the negative samples as follows: we select 10 negative segments form all segments of training data. All segments contain positively selected by other users and unobserved segments of the training video. This negative sampling technique encourages more negative segments appeared in user-item graph with user ratings.  The reason is that, if a segment is never rated by any user, it is an isolated point of this graph without any neighbor, and could be learned with higher-order graph structure.  In the model learning process, we use
Adam optimizer with a learning rate of 0.001 and a mini-batch size of 50. In our TransGRec framework, we choose the embedding dimension D in the set $[32,64,128,256]$, and find {\small$D=128$} reaches the best performance.
Besides, the regularization parameter $\lambda$ is set in range $[0.1,1,10]$, and $\lambda =1$ reaches the best performance and are is stable. For each propagation layer as shown in Eq.\eqref{eq:gcn_u} and Eq.\eqref{eq:gcn_i}, we use ReLU function as the activation function to transform the output embedding, and use the mean pooling as the pooling operation. For transfer network~$T$, we choose to use a two-layer MLP. For discriminator~($D$) in transfer network~$T$, we train a two-layer MLP. There are some parameters in the baselines, we tune all these parameters to ensure these baselines reach the best performance.

\begin{table*}
 \begin{small}
  \centering
  \vspace{-5pt}
  \setlength{\belowcaptionskip}{-3pt}
  \caption{Effects of different propagation layer depth K in TransGRec-E.} \label{tab:propagation_depth}
  \vspace{-8pt}
  \begin{center}
     \scalebox{1}{
  \begin{tabular}{|c|c|c|c|c|c|c|c|c|c|c|}
      \hline
            Depth K&MAP$\uparrow$&Improve&NMSD$\downarrow$&Improve&HR@5$\uparrow$&Improve&NDCG@5$\uparrow$&Improve&Recall@5$\uparrow$&Improve\\
            \hline
            \textbf{K=2}&\textbf{0.3113}&-&\textbf{0.3702}&-&\textbf{0.3066}&-&\textbf{0.2687}&-&\textbf{0.2873}&-\\
            \hline
            K=0&0.3020&-2.99\%&0.3921&-5.92\%&0.2834&-7.57\%&0.2542&-5.40\%&0.2617&-8.91\% \\
            \hline
            K=1&0.3073&-1.28\%&0.3816&-3.08\%&0.2999&-2.19\%&0.2617&-2.61\%&0.2802&-2.47\% \\
            \hline
            K=3&0.2990&-3.95\%&0.3989&-7.75\%&0.2941&-4.08\%&0.2553&-4.99\%&0.2741&-4.59\% \\
      \hline
  \end{tabular}}
  \end{center}
 \end{small}
\end{table*}

\begin{table*}
 \begin{small}
  \centering
  \setlength{\belowcaptionskip}{-3pt}
  \caption{Effects of different propagation layer depth K in TransGRec-A.} \label{tab:propagation_depth_a}
  \vspace{-8pt}
  \begin{center}
     \scalebox{1}{
  \begin{tabular}{|c|c|c|c|c|c|c|c|c|c|c|}
      \hline
            Depth K&MAP$\uparrow$&Improve&NMSD$\downarrow$&Improve&HR@5$\uparrow$&Improve&NDCG@5$\uparrow$&Improve&Recall@5$\uparrow$&Improve\\
            \hline
            \textbf{K=1}&\textbf{0.3174}&-&\textbf{0.3769}&-&\textbf{0.3153}&-&\textbf{0.2779}&-&\textbf{0.2940}&-\\
            \hline
            K=0&0.3030&-4.54\%&0.3989&-5.84\%&0.2861&-9.26\%&0.2586&-6.94\%&0.2633&-10.44\% \\
            \hline
            K=2&0.3091&-2.61\%&0.3854&-2.26\%&0.2911&-7.68\%&0.2612&-6.01\%&0.2723&-7.38\% \\
      \hline
  \end{tabular}}
  \end{center}
 \end{small}
 \vspace{-10pt}
\end{table*}

\subsection{Overall Comparison}
We show the evaluation results of various metrics in Table~\ref{tab:overall comparision}. Among all the baselines, Video2GIF is a generalized video highlight recommendation model that presents the same video highlights to all users. All the remaining models show better performance than Video2GIF, indicating the soundness of personalization. DropoutNet and CDL improve over PHD-GIFs and  SVD-M about 20\% with more precise user embedding. CDL further improves DropoutNet. We guess a possible reason is that, in order to adapt the problem to the new user setting, DropoutNet optimizes the original rating based loss and the new user dropout loss at the same time. Thus, it does not show the best performance under the scenario without new users. As LapReg is implemented with an additional Laplacian regularization term on top of the remaining best baseline~(i.e., CDL), LapReg shows better performance compared to CDL on most evaluation metrics.

Our proposed two detailed models of the TransGRec framework consistently outperform the best baselines, showing the effectiveness of better user modeling through user embedding propagation layers in the user-item graph. On average, TransGRec-A improves the best baseline about 13\% in MAP, 1\% in NMSD, 16\% in HR@5, 20\% in NDCG@5 and 24\% in Recall@5. When comparing TransGRec-A and TransGRec-E, we find on average TransGRec-A shows better performance than TransGRec-E. We guess a possible reason is that TransGRec-A could measure the difference of two complex distributions in non-Euclidean space.

In order to better evaluate the performance of recommendation task, we report the Top-N ranking performance with different $N$ values about various models. The detailed experiment results are shown in Table \ref{tab:hr_ndcg_recall_topk}. We also find the similar observations as Table \ref{tab:overall comparision}, with our proposed TransGRec always shows the best performance under various Top-N values. Based on the overall experimental results, we could empirically conclude that our proposed TransGRec framework outperforms all the baselines under both video detection metrics and ranking metrics.

\begin{figure}[htb]
  \begin{center}
  \vspace{-0.3cm}
      \subfigure{\includegraphics[width=42mm]{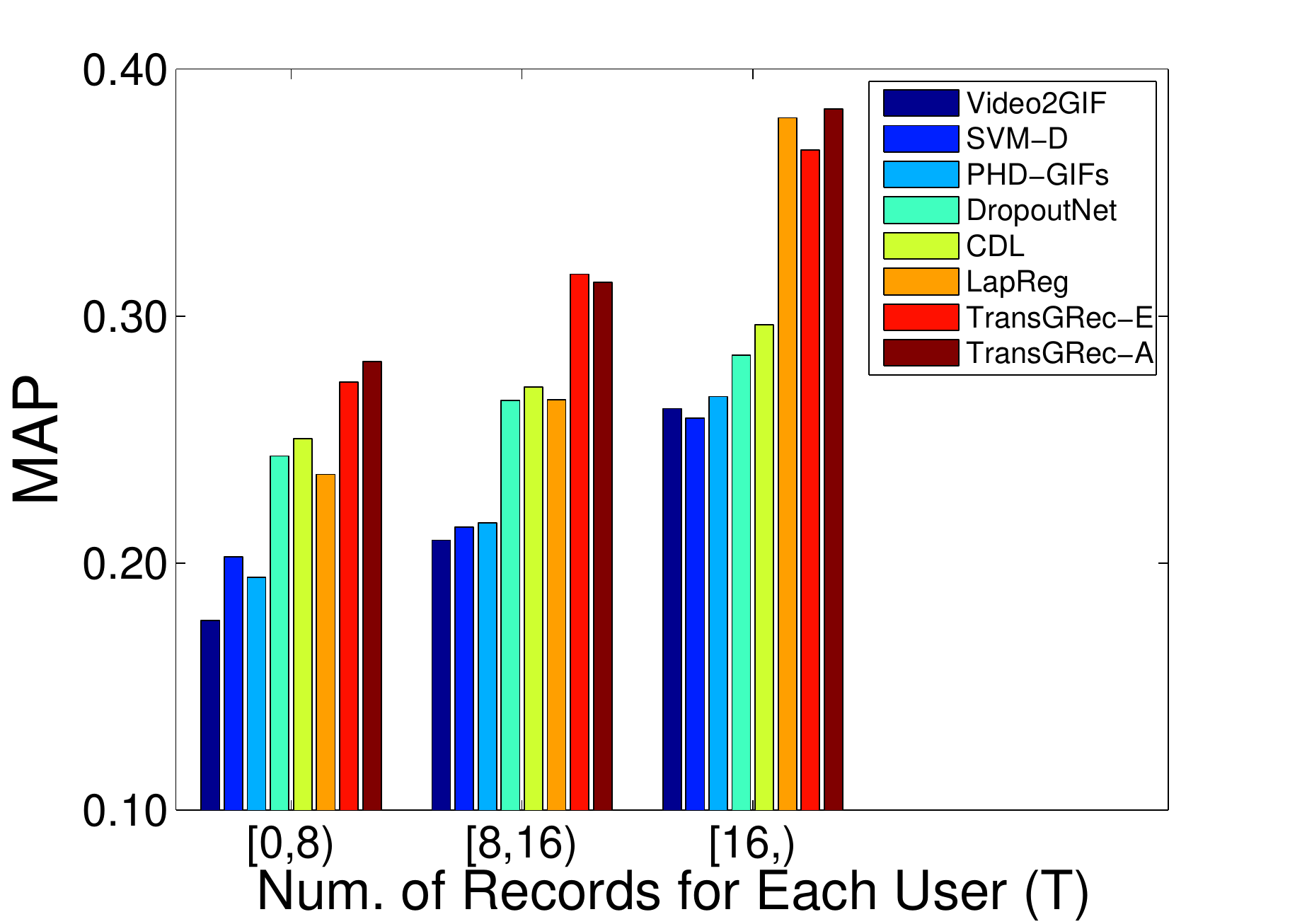}}
      \subfigure{\includegraphics[width=42mm]{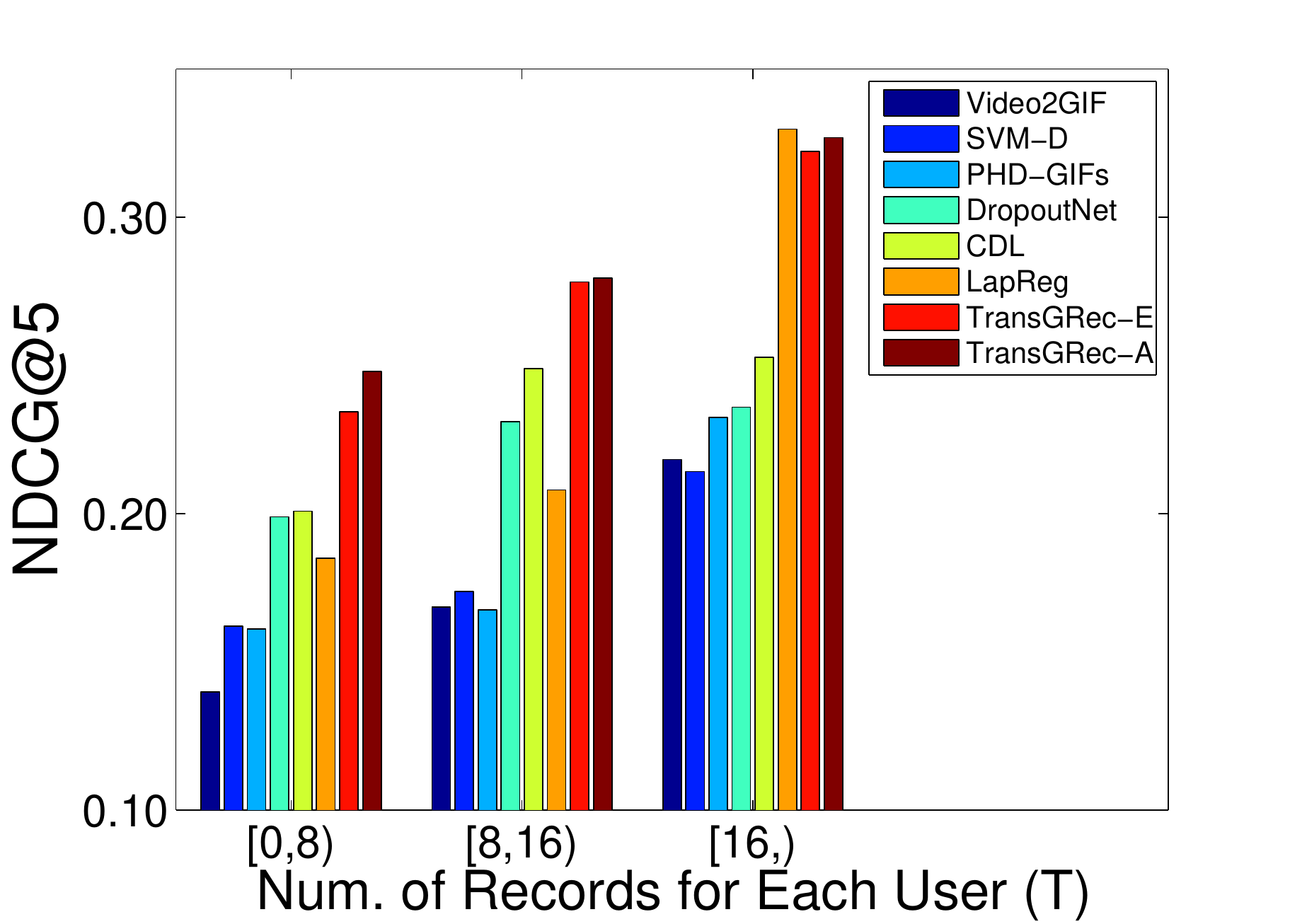}}
  \end{center}
  \vspace{-0.3cm}
  \caption{Performance under different data sparsity.} \label{fig:sparsity_results}
  \vspace{-0.3cm}
\end{figure}

\begin{figure}[htb]
  \begin{center}
      \subfigure{\includegraphics[width=80mm]{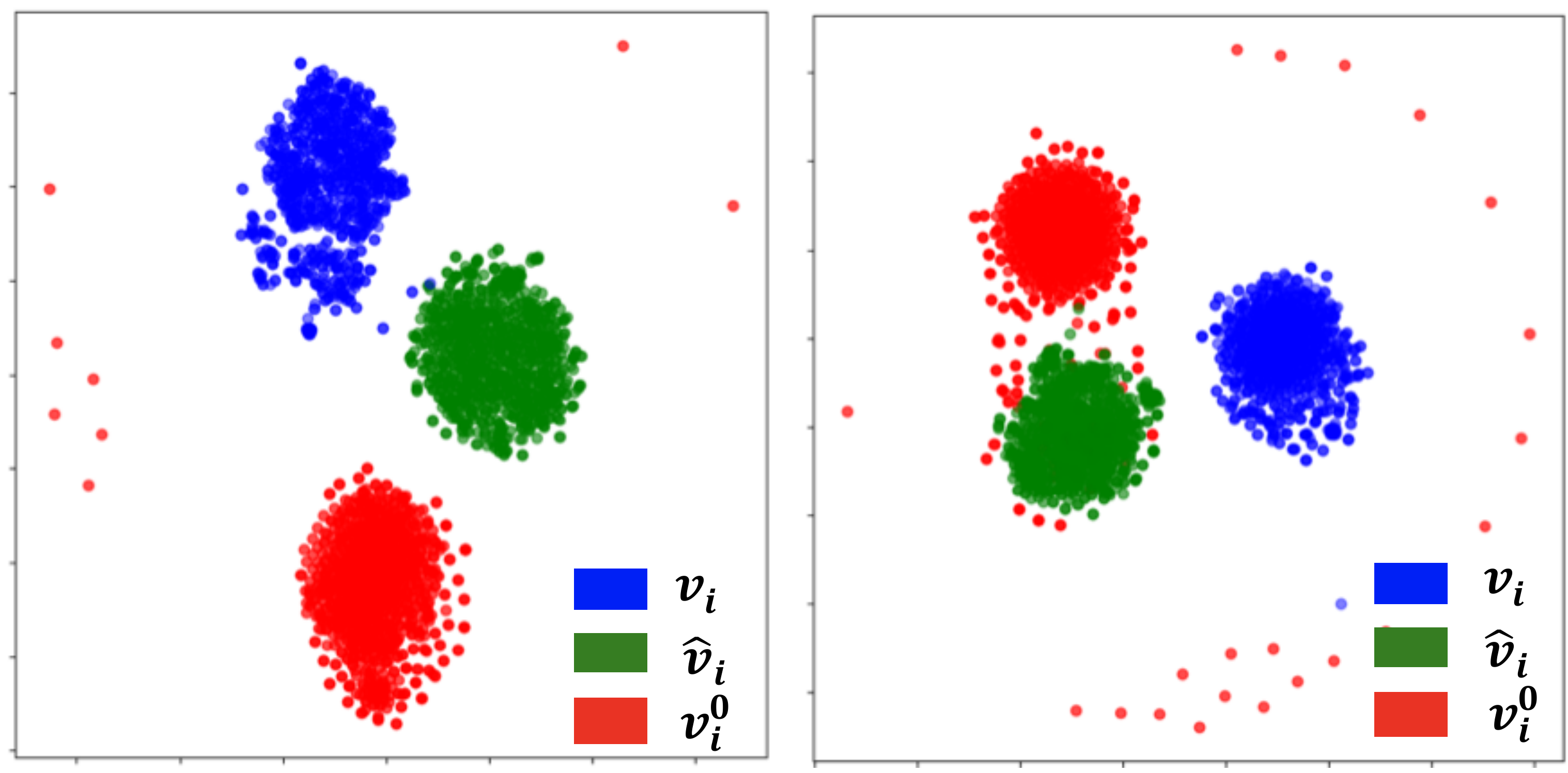}}
  \end{center}
  \vspace{-0.2cm}
  \caption{The t-SNE visualization of three types of item embeddings of different models. Color of nodes indicates the type of node embeddings. Red: `` the layer-0 embedding in embedding propagation layers~($\mathbf{v}^0_i$)'', Blue:``final item embedding output by the embedding propagation layers~($\mathbf{v}_i$)'', green:``approximated item embedding learned by the transfer network~($\hat{\mathbf{v}}_i$)''.  The left figure shows the results of TransGRec-E, and the right figure shows the results of TranGRec-A.} \label{fig:tsne_results}
  \vspace{-0.3cm}
\end{figure}

\subsection{Performance under Different Data Sparsity}
We split users into different groups and observe the performance under different sparsity. Specifically, we split all users into three groups based on the number of the rated items in the training data, and the performance of all models under different sparsity are shown in Figure \ref{fig:sparsity_results}. Due to page limit, we do not show the results of NMSD as it shares similar trends as MAP. Also, the results of HR@5 and Recall@5 are similar to NDCG@5.
E.g,  for each user in the group of $[8,16)$,  the number of her training records satisfies $8\leq |\mathbf R_{a}| \textless 16$. As illustrated in this figure, with the increase of the user interaction records, the performance increases quickly for all models, as all models need to rely on user rated items for embedding learning. Our proposed models consistently outperform the baselines under all user groups. E.g., TransGRec-A improves over the best baselines  about 23\% on the $[0,8)$ user group with the NDCG@5 metric. All baselines show similar performance trend, expect the LapReg model, which shows similar performance
as TransGRec on the $[16,)$ group. However, LapReg does not perform well on the sparser user groups. We guess a possible reason is that, LapReg relies on the local link structure for graph regularization. When users have limited links, the regularization term is not reliable with limited first-order neighbors. As users have more rating records, the performance of LapReg increases. In contrast, TransGRec could better model the higher-order graph structure, and it shows relatively higher performance even with cold-start users.

\subsection{Detailed Model Analysis}
In this part, we analyze the impact of different propagation layer in depth $K$. Table~\ref{tab:propagation_depth} and Table~\ref{tab:propagation_depth_a} summarize the results of two models with different $K$ values. The column of ``Improve'' shows the performance changes compared to the best setting of model, i.e., {\small$K$=2}. When the propagation layer depth {\small$K$=0}, TransGRec only uses initialized user embedding and not aggregates neighbor feature. As shown in Table~\ref{tab:propagation_depth}, when the propagation depth increases from {\small$K$=1} to {\small$K$=2}, the performance improves and $K$=2 reaches the best performance. However, the performance drops when $K$=3 as three layers may introduce noisy neighbors of the graph. The performance drops when $K$=3 has also been observed in GCN based models~\cite{ICLR2017graph,sigir2019diffnet}. The same phenomenon also occurs in Table~\ref{tab:propagation_depth_a}, the difference is that the best performance reaches when {\small$K$=1} for TransGRec-A.

A key characteristic of our proposed TransGRec module is  designing a transfer network to approximate the learned embeddings from graph neural networks. Therefore, it is natural to ask: does the transfer network show ability to mimic the graph network embeddings. To answer this question, for both methods of TransGRec-E and TransGRec-A, after the training process converges, we visualize three kinds of item embeddings:   the layer-0 embedding in the embedding propagation layers~($\mathbf{v}^0_i$) of the graph neural network,  the final item embedding output by the embedding propagation layers~($\mathbf{v}_i$), and the approximated item embedding learned by the transfer network~($\hat{\mathbf{v}}_i$). We randomly select 1000 items and map these three categories of item embeddings into a 2-D space with the t-SNE package~\cite{maaten2008visualizing} . The results are shown in Figure~\ref{fig:tsne_results}. Visualizations of the two models show similar phenomenon. The representation exhibits clustering effects of different kinds of item embeddings. Specifically, the distance between the fused item embedding and graph embedding output is large, showing that the graph neural network could transform the fused item embedding into a different space through iterative graph embedding propagation layers. We observe the relative distance between the transferred embeddings and the initial embeddings,  is smaller than that between the the graph output embeddings and the initial embeddings. This phenomenon shows that the transfer network could partially mimic the graph embedding function to learn useful transfer embedding, and validates the effectiveness of the transfer network. However, there are still a gap between the transferred approximated embeddings and the graph embedding output. We guess the reason is that, the graph structure is very complex, while relying on a transfer network could not well capture all the information hidden in the graph.  We would leave the problem of how to better design a transfer network that well captures the graph structure as our future work.

\section{Related Works}

\subsection{Video Highlight and Personalization}
Automatic video  highlight extraction and summarization deals with selecting representative segments from videos~\cite{CVPR2014real,CVPR2015video}. These models learned interesting, representative, or diversified  visual content segments based on well-defined optimization goals~\cite{CVPR2014real,ECCV2016}. As many users edited videos in online platforms, some researchers proposed to leverage the wisdom of the crowds as ground-truth or priors to guide video highlight extraction~\cite{CVPR2013prior,CVPR2015video,CVPR2016video2gif}. However, these proposed models neglected users' personalized preferences. With the huge boom of video editing platforms and APPs, recently researchers published a personalized video highlight dataset that records each user's liked segments of videos, and  proposed a personalized highlight detection model~\cite{MM2018phd}. The personalization is achieved by adapting the input of each user as an aggregation of her liked segments of videos, and the proposed personalization model showed better performance compared to the state-of-the-art generalized highlight extraction model~\cite{CVPR2016video2gif}, indicating the soundness of personalization in video highlight recommendation. However, as each user's annotated records are limited, the recommendation performance is still far from satisfaction.

\subsection{Classical Recommendation Models}
Given user-item interaction behavior, CF achieves high performance by learning users and items in a low dimensional collaborative space~\cite{BPR2009,lian2020personalized,lian2020lightrec}. However,
CF  models fail to tackle the new item problem as these models relied on the user-item interaction behavior for recommendation. With the huge success of deep learning in computer vision and related areas, many state-of-the-art content based models learn to align users and items in a semantic content space with deep learning techniques~\cite{CVPR2016CDL,ICCV2017videorec,IJCAI2019semantic}. Some hybrid recommendation models were proposed to leverage both user behavior data and item content for recommendation~\cite{AAAI2016vbpr,NIPS2017dropoutnet,KDD2015CDL,TKDE2019hierarchical}. However, most of these models could not tackle the new item problem, as each item representation is a hybrid representation with both the collaborative information and the semantic content representation~\cite{AAAI2016vbpr}. There are two related hybrid recommendation works that could adapt to new item issue~\cite{NIPS2017dropoutnet,KDD2015CDL}. In the proposed Collaborative Deep Learning model, when the item does not appear in the training data, the item embedding degenerated to the semantic embedding without any collaborative signal~\cite{KDD2015CDL}.  Researchers proposed a DropoutNet to address the cold start problem in recommender systems~\cite{NIPS2017dropoutnet}. By treating cold start as the missing preference data, DropoutNet modified the learning procedure with dropout techniques to explicitly condition the model for missing inputs. Besides the application scenario, we also differ greatly from these works as we reformulate the hybrid recommendation with new items from a graph perspective. Therefore, our proposed model could better capture the relationships between users and items for recommendation, especially for users who have limited records.

\subsection{Graph Learning Models and Applications in Recommendation}
Recently, Graph Convolutional Networks~(GCN) have shown huge success for graph representation learning~\cite{ICLR2017graph,ICLR2017semi}. These models generate node embeddings in a message passing or information propagation manner in the graph, with a node embedding is recursively computed by aggregating neighboring nodes' information. In fact, researches have already shown that GCNs could be seen as a special kind of Graph Laplacian smoothing~\cite{li2018deeper,ICLR2019predict}. With the requirement of quickly generating embeddings for unseen nodes, GraphSAGE provides an inductive learning approach that
learns a mapping function from node features and node links to node embedding~\cite{NIPS2017inductive}.
Due to the huge success of the GCNs, several models have attempted to utilize the idea of GCNs for recommendation, such as user-item bipartite graph in collaborative filtering~\cite{kdir18,sigir2019ngcf}, user influence diffusion in social recommendation~\cite{sigir2019diffnet}, and item-item similarity graph for similar item recommendation~\cite{KDD2018graph}. Most of these GCN based recommendation models are based on CF, and could not tackle the new item problem~\cite{kdir18,sigir2019ngcf,sigir2019diffnet}. Specifically, PinSage is one of the state-of-the-art inductive content based GCN~\cite{KDD2018graph}. By taking both item-item correlation graph and item features as input, PinSage learned transformed item embedding with graph convolutional operations. However, in the real-world, new items do not have user ratings, i.e., the node neighbors for embedding learning. Our main technical contribution lies in transferring the knowledge learned from the graph neural networks, such that our proposed TransGRec framework is also applicable to the new item without any link information.

\section{Conclusion}
In this paper, we designed a TransGRec framework for personalized video highlight recommendation.  We based TransGRec on an graph neural network model, and proposed to propagate user embeddings in the graph to alleviate the cold-start user problem. We further proposed a transfer network that learns to transform the item content embedding to the graph neural network space. Therefore, TransGRec is an inductive graph based recommendation approach. Please note that, though we use the personalized video recommendation as an application scenario, our proposed model is generally applicable to any content based recommendation tasks. Finally, extensive experimental results on a real-world dataset clearly showed the effectiveness of our proposed model under various evaluation metrics. In the future, we would like to  explore how to design more sophisticated transfer network for graph embedding approximation. Besides, we would like to apply and validate the effectiveness of our proposed framework in the general recommender systems.

\section*{Acknowledgements}

This work was supported in part by the National Natural Science Foundation of China(Grant No.61725203, 61972125, U1936219, 61722204, 61932009 and 61732008).

\bibliographystyle{ACM-Reference-Format}
\bibliography{VISGCN}

\end{document}